\documentstyle[epsfig,psfig]{texas}

\begin{document}

\title{MAXIMUM REDSHIFT AND MINIMUM ROTATION PERIOD OF NEUTRON STARS}
\author{P. Haensel$^{1,2}$, J.P. Lasota$^{3}$, and J.L. Zdunik$^{2}$}

\address{(1) Departement d'Astrophysique Relativiste et de Cosmologie\\
UPR 176 du CNRS, Observatoire de Paris, Section de Meudon,
 F-92195 Meudon Cedex, France\\
}
\address{(2)N. Copernicus Astronomical Center, Polish
           Academy of Sciences\\
Bartycka 18, PL-00-716 Warszawa, Poland\\
}
\address{(3) Istitut d'Astrophysique de Paris, CNRS\\ 
98bis Boulevard Arago, F-78014 Paris, France\\
{\rm Email: jlz@camk.edu.pl, haensel@camk.edu.pl, 
lasota@iap.fr}}

\begin{abstract}
 The lower bound on the period of uniform rotation of neutron
stars (NS) with causal equation of state (EOS), $P_{\rm min}$, is shown
to be determined, with a high precision, by the upper bound on the
surface redshift of static neutron star models, $z_{\rm max}$. We
obtain  $z_{\rm max}=0.851$, which turns out to be about $5\%$ lower
than upper bound obtained by Lindblom (1984). Using then the empirical
formula for the maximum rotation frequency of uniform rotation of
neutron star models with realistic causal  EOS, derived by Lasota et
al. (1996), we establish an approximate  relation between $z_{\rm max}$
and $P_{\rm min}$,  in which  $P_{\rm
min}$  is proportional to the maximum mass of static NS, $M_{\rm
s}^{\rm max}$. In this way we reproduce, within 2\%,  the  exact
formula of Koranda et al. (1997) based on extensive exact numerical
calculations. Replacing  $M_{\rm s}^{\rm max}$ by the maximum measured
NS mass, $1.442~M_\odot$, we get a lower bound $P_{\rm min}\simeq
0.288$~ms. \end{abstract}

\section{Introduction }
The lower limit on the period of a uniformly rotating neutron star depends  
on the - largely unknown -  equation of state (EOS) of dense 
matter above the
nuclear density.  Therefore, an  uncertainty in the high density EOS
implies a large uncertainty in the minimum period of uniform rotation,
$P_{\rm min}$ (see, e.g. \cite{FI87}, \cite{FIP89}, \cite{Salgado1}, 
 \cite{Salgado2}, \cite{CookST94}). 
  Hence, it is,
therefore, of interest to find a lower limit on $P_{\rm min}$, that is
independent of the EOS. This limit results from the condition of
causality, combined with the requirement that EOS yields neutron stars
with masses compatible with observed  ones (currently the highest
accurately measured neutron star mass is $M_{\rm obs}^{\rm
max}=1.442~{\rm M}_\odot$ \cite{TaylorW89}).  It will be
hereafter referred to as $P_{\rm min}^{\rm CL}$.

The first calculation of $P_{\rm min}^{\rm CL}$ was done by Glendenning
\cite{Glend92}, who found the value of $0.33$~ms. Glendenning 
\cite{Glend92}, however,
used a rather imprecise empirical formula, to calculate a lowest
$P_{\rm min}$ by using the parameters (mass and radius) of the maximum
mass configurations of a  family of non-rotating neutron star models.
His result, therefore, should be considered only as an estimate of
$P_{\rm min}^{\rm CL}$. Recently, Koranda et al. \cite{Koranda97} extracted the
value of $P_{\rm min}^{\rm CL}$ from extensive {\it exact} calculations
of uniformly rotating neutron star models. They have shown, that the
method of Glendenning \cite{Glend92} overestimated the value of $P_{\rm
min}^{\rm CL}$ by 6\%. The result of  Koranda et al. \cite{Koranda97}  
calculations  can be summarized  in a formula
\begin{equation}
P_{\rm min}^{\rm CL} = 
0.196~{M_{\rm obs}^{\rm max}\over {\rm M}_\odot}~{\rm ms}~, 
\label{Koranda.Pmin}
\end{equation}
which combined with measured mass of PSR B1913+16 yields 
today's lower bound for $P_{\rm min}^{\rm CL}=0.282$ ms. 
This absolute bound on the 
minimum period was  obtained for  the  ``causality limit
(CL)  EOS''  
$p=(\rho-\rho_{\rm 0})c^2$, which yields neutron star models of the surface 
density $\rho_{\rm 0}$ and is maximally stiff (${\rm d}p/{\rm d}\rho = c^2$)  
everywhere within the star; it does not depend on the value of 
$\rho_0$. In the present  contribution we show that Eq. (1) 
can be reproduced  using an empirical  formula for $P_{\rm min}$ 
derived for 
{\it realistic} causal EOS, 
 combined  with an  upper bound on the relativistic surface redshift  
 for static neutron stars with 
causal EOS. 
%
\section{Relation  between $x_{\rm s}$ and $P_{\rm min}$}
Numerical results for maximum frequency of stable uniform rotation of 
neutron stars, 
obtained by Salgado et al. \cite{Salgado1}, \cite{Salgado2} 
 for a broad set of realistic, causal 
EOS of dense matter, can be  reproduced within better than 5\%  by an empirical 
formula \cite{HSB95} 
\begin{equation}
{\rm simple~empirical~formula:~~~~~~~~~}
\Omega_{\rm max}\simeq 0.673 \left({G M_{\rm s}\over R_{\rm s}^3}\right)
^{1/2}~{\rm s^{-1}}~,
\label{Omega:e.f.1}
\end{equation}
where $M_{\rm s}$ and $R_{\rm s}$ are, respectively, 
 mass and radius of {\it static} 
 configuration with maximum allowable mass. Numerical prefactor in 
Eq.(\ref{Omega:e.f.1}) is independent of the EOS from the set considered 
in \cite{Salgado1}, \cite{Salgado2}. 

Defining a relativistic compactness parameter 
\begin{equation}
x_{\rm s}\equiv {2GM_{\rm s}\over R_{\rm s}c^2}~,
\label{xs.def}
\end{equation}
related to the surface redshift of static configuration with maximum allowable
mass by 
\begin{equation}
z_{\rm s}=\left(1-x_{\rm s}\right)^{-{1/2}}-1~,
\label{zs.xs}
\end{equation}
we can transform Eq. (\ref{Omega:e.f.1})  into an approximate empirical 
formula, expressing the minimum period of rotation of neutron star, 
 $P_{\rm min}=2\pi/\Omega_{\rm max}$, in terms of  $M_{\rm s}$ and $x_{\rm s}$, 
\begin{equation}
{\rm simple~empirical~formula ~~\Longrightarrow ~~~}
P_{\rm min}\simeq {0.130\over x_{\rm s}^{3/2}}
{M_{\rm s}\over {\rm M}_\odot}~{\rm ms}~. 
\label{Pmin:e.f.1}
\end{equation}
Clearly, for a given maximum mass of static configurations, $M_{\rm s}$, 
the lower bound on  $P_{\rm min}$ will be obtained by replacing $x_{\rm s}$ 
by its upper bound, $x_{\rm s,max}$.

The formula for $P_{\rm min}$ can be improved by using more precise empirical
formula for $\Omega_{\rm max}$.  
Such a formula was constructed  by Lasota et al. \cite{LasotaHA96}. 
As shown in \cite{LasotaHA96}, 
numerical results  of \cite{Salgado1}, \cite{Salgado2}  
 for the maximum frequency 
of uniform stable rotation can be reproduced (within better than 2\%),   
for a broad set of realistic causal EOS of dense 
matter, by an improved empirical formula
\begin{equation}
{\rm improved ~empirical~ formula:}~~~~~~~~~
\Omega_{\rm max}
 \simeq  {\cal C}(x_{\rm s})
\left(
{G M_{\rm s}\over R_{\rm s}^3}
\right)^{1/2}~{\rm s^{-1}},
\label{Omega:e.f.2}
\end{equation}
\vskip 2mm
where ${\cal C}(x_{\rm s})$ is 
a universal (i.e. independent of the EOS) function  of the compactness 
parameter 
$x_{\rm s}$, 
\begin{equation}
{\cal C}(x_{\rm s})= 
0.468 + 0.378 x_{\rm s}~. 
\label{C}
\end{equation}
Combining Eq. (\ref{Omega:e.f.2}) and Eq. (\ref{C}) we get
\vskip 2mm 
\begin{equation}
{\rm improved~ empirical~ formula~~\Longrightarrow ~~~} P_{\rm min}
\simeq  
{8.754\times 10^{-2}\over 
{\cal C}(x_{\rm s})x_{\rm s}^{3/2}} 
~{M_{\rm s}\over {\rm M}_\odot}~{\rm ms}.~~~
\label{Pmin:e.f.2}
\end{equation}
\vskip 2mm
Both  ``empirical formulae'' for $P_{\rm min}$,  
Eq.(\ref{Pmin:e.f.1}) and Eq.({\ref{Pmin:e.f.2}),  imply, 
that at given maximum mass of a spherical  configuration, the  
minimum  rotation period is obtained for the maximum 
value of $x_{\rm s}$.   At fixed $x_{\rm s}$, the value of $P_{\rm min}$ 
is proportional to $M_{\rm s}$. Neutron 
stars for which masses have been measured,  rotate so slowly
that their structure  
can be very well approximated by that of a spherical star. Observations impose 
thus a condition $M_{\rm s}\ge M^{\rm max}_{\rm obs}$. 
\section{Upper bounds on surface redshift}
To  minimize $P_{\rm
min}$ for given $M_{\rm obs}^{\rm max}$,  we have to look for an EOS
which yields maximum $x_{\rm s}$ at $M_{\rm s}=M_{\rm obs}^{\rm max}$.
It is well known, that if one relaxes the condition of causality, the
absolute upper bound on $x_{\rm s}$ for stable neutron star models is
reached for an incompressible fluid  (i.e., $\rho=const.$) EOS; the
value of $x_{\rm s}$ is then independent of $M_{\rm s}$ and equal $8/9$; 
it corresponds to an upper bound on $z_{\rm s}$, equal 2 
(see, e.g., \cite{ST83}).  

 It is therefore rather natural
to expect that in order   to maximize $z_{\rm s}$  under the condition
of causality, one has to maximize the sound velocity throughout the star.
Together with condition of density continuity in the stellar interior
this  points out at the CL EOS, $p=(\rho-\rho_0)c^2$,  as to that which
yields ``maximally compact neutron stars''; 
introducing density
discontinuities does not increase the value of $z_{\rm s}$, see
\cite{GZ95}.  
 Note, that the value of $z_{\rm s}$ for CL EOS does
not depend on $\rho_0$ (and therefore is $M_{\rm s}$-independent). It
represents an absolute upper bound on $z_s$ for causal EOS, $z_{\rm
max}$.  Our numerical calculation gives $z_{\rm s}({\rm
CL~EOS})=z_{\rm max}= 0.8509$. 
This corresponds to an absolute upper bound on the compactness parameter 
of neutron star models with causal EOS, 
$x_{\rm max}= 0.7081$. 

Let us consider the effect of the presence of a crust (more generally,
of an envelope of normal neutron star matter). For a given EOS of the
normal envelope, the relevant (small) parameter is the ratio $p_{\rm
b}/\rho_{\rm b}c^2$, where $p_{\rm b}$ and $\rho_{\rm b}$ are,
respectively, pressure and mass density at the bottom of the crust
 \cite{Lindblom84}.  The case of $p_{\rm b}=0$ corresponds to stellar
models with no normal crust.  Numerical calculations show, that adding
a crust onto a CL EOS core implies an increase of $R_{\rm s}$, which is
linear in $p_{\rm b}/\rho_{\rm b}c^2 $; for a solid crust we have
typically $p_{\rm b}/\rho_{\rm b}c^2 \sim 10^{-2}$.  The change
(increase) in $M_{\rm s}$ is negligibly small; it turns out to be
quadratic in $p_{\rm b}/\rho_{\rm b}c^2$. This implies, that the
decrease of $x_{\rm s,max}$, and of the maximum surface redshift $z_{\rm
s,max}$, due to the presence of a crust,
is  proportional to $p_{\rm b}/\rho_{\rm b}c^2$. This   is consistent
with Table 1 of Lindblom \cite{Lindblom84}. However, the extrapolation of his
results to $p_{\rm b}=0$ yields $z_{\rm max}=0.891$, which is nearly
5\% higher than our value of $z_{\rm max}$ ! This might reflect a lack
of precision of the variational method used by Lindblom 
\cite{Lindblom84}, which
led to an overestimate of the value of $z_{\rm max}$.  It should be
stressed that while a precise determination of $M_{\rm max}\equiv
M_{\rm s}$ for static neutron star models is rather easy, determination
of the precise value of the radius of the maximum mass
configuration, $R_{\rm s}$, (with the same relative precision as
$M_{\rm s}$) and consequently of the value of $x_{\rm s}$ (with, say,
four significant digits), is much more difficult and requires a rather
high precision of numerical integration of the TOV equations.
\section{Lower bound on $P_{\rm min}$}
We restrict ourselves to the case of the absolute lower  
bound on $P_{\rm min}$, obtained for neutron star models with no
crust.   
Inserting the value of $x_{\rm s,max}$ into  Eq. (\ref{Pmin:e.f.1}) we get 
\begin{equation}
{\rm simple~ empirical~ formula}~~\Longrightarrow~~~
P^{\rm CL}_{\rm min}
\simeq 0.22 { M_{\rm obs}^{\rm max}\over {\rm M}_\odot}~{\rm ms}~.
\label{bound.emp1}
\end{equation}
The current lower bound on $P$, resulting from the above equation, is 0.32 ms, 
which should be compared with the 
lower bound of 0.282 ms, obtained in extensive 
exact numerical calculations by Koranda et al. \cite{Koranda97}.

Using the improved empirical formula, Eq. (\ref{Pmin:e.f.2}), we get
\begin{equation}
{\rm improved~ empirical~ formula}~~\Longrightarrow~~~
P^{\rm CL}_{\rm min}
\simeq 0.1997 { M_{\rm obs}^{\rm max}\over {\rm M}_\odot}~{\rm ms}~, 
\label{bound.emp2}
\end{equation}
which implies 
 a current lower bound of     
$0.288$~ms,   only 2\% higher than the 
exact result of Koranda et al. \cite{Koranda97}.

The formula (\ref{bound.emp2}) deserves an additional comment.  In
numerical calculations, of  a family of stable uniformly rotating
stellar models, for a given EOS of dense matter, one has to distinguish
between the rotating configuration of maximum mass, which corresponds
to the rotation frequency  $\Omega_{M_{\rm max}}({\rm EOS})$, and the
maximally rotating one, which rotates at $\Omega_{\rm max}({\rm EOS})$
 \cite{CookST94}, \cite{StergF95}. Notice, that
determination of a maximum mass rotating configuration (and therefore
of $\Omega_{M_{\rm max}}$)  is a much  simpler task than the
calculation of exact value of $\Omega_{\rm max}$, which is time
consuming and very demanding as far as the precision of numerical
calculations is concerned. Usually,  both configurations are very close
to each other, and $\Omega_{\rm max}$ is typically only 1-2\% higher
than  $\Omega_{M_{\rm max}}$; such a small difference is within the
typical precision of the empirical formulae for $\Omega_{\rm max}$.
Actually, the formula for ${\cal C}(x_{\rm s})$, Eq. (\ref{Omega:e.f.2}),  
was fitted
to the values of $\Omega_{M_{\rm max}}({\rm EOS})$ calculated in
 \cite{Salgado1,Salgado2}. Therefore, Eq.(\ref{bound.emp2}) should in
principle be used to evaluate the causal lower bound to $P_{{\rm
min},M_{\rm max}}$; it actually   reproduces, within 0.2\%,  the exact
formula for this quantity,   obtained by Koranda et al. \cite{Koranda97} [see
 their Eq. (8)].

General experience shows that -  in contrast to interpolation -
extrapolation is a risky procedure. The fact  
that in our case extrapolation of an empirical formula
yields  - within 2\% -  the value of $P_{\rm min}$ 
 of Koranda et al. \cite{Koranda97} (and reproduces their value of 
$P_{{\rm min},M_{\rm max}}$), proves the usefulness of 
compact ``empirical expressions'' which might summarize, in a
quantitative way,   a relevant content  of extensive 
numerical calculations  of  uniformly rotating neutron star models.  
\section*{Acknowledgments}
This research was partially supported  by the KBN grant No. 2P03D.014.13. 
 During his stay at DARC, P. Haensel was supported by the PAST Professorship  
of French MENESRT. 

\section*{References}

\par

\end{document}